\documentclass[conference]{IEEEtran}
\IEEEoverridecommandlockouts

\usepackage{cite}
\usepackage{caption}
\usepackage{subcaption}
\usepackage[implicit=false,bookmarks=false]{hyperref}
\usepackage{amsmath,amssymb,amsfonts}
\usepackage{algorithmic}
\usepackage{graphicx}
\usepackage{textcomp}
\usepackage{xcolor}
\usepackage{mathtools}
\usepackage{booktabs}
\usepackage{siunitx}
\usepackage{multirow}
\usepackage[none]{hyphenat}
\usepackage{lipsum}
\usepackage{multicol}
\usepackage{stfloats}

\usepackage[export]{adjustbox}
\usepackage{balance}

\begin{document}
\title{RVE-CV2X: A Scalable Emulation Framework for Real-Time Evaluation of CV2X-based Connected Vehicle Applications}
\author{Ghayoor Shah$^*$, Md Saifuddin$^*$, Yaser P. Fallah$^*$, Somak Datta Gupta$^\dagger$\\
$^*$Connected and Autonomous Vehicle Research Lab (CAVREL),\\
University of Central Florida, Orlando, FL\\
$^\dagger$Ford Motor Company, Dearborn, MI\\ 
\{gshah8, md.saif\}@knights.ucf.edu, yaser.fallah@ucf.edu, sdattagu@ford.com
        }

\maketitle

\begin{abstract}
Vehicle-to-Everything (V2X) communication has become an integral component of Intelligent Transportation Systems (ITS) due to its ability to connect vehicles, pedestrians, infrastructure, and create situational awareness among vehicles. 
Cellular-Vehicle-to-Everything (C-V2X), based on 3rd Generation Partnership Project (3GPP) Release 14, is one such communication technology that has recently gained significant attention to cater the needs of V2X communication. 
However, for a successful deployment of C-V2X, it is of paramount significance to thoroughly test the performance of this technology. It is unfeasible to physically conduct a V2X communication experiment to test the performance of C-V2X by arranging hundreds of real vehicles and their transceiving on-board units. Although multiple simulators based on frameworks such as NS-3, OMNET++ and OPNET have proven to be reliable and economic alternatives to using real vehicles, all these simulators are time-consuming and require several orders of magnitudes longer than the actual simulation time. As opposed to physical field- and simulation-based testing, network emulators can provide more realistic and repeatable results for testing vehicular communication. This paper proposes a real-time, high-fidelity, hardware-in-the-loop network emulator (RVE-CV2X) based on C-V2X mode 4 that can provide scalable, reliable and repeatable testing scenarios for V2X communication. The accuracy of this emulator is verified by comparing it to an already validated C-V2X simulator based on the NS-3 framework.

\end{abstract}
\section{Introduction}
Connected and automated vehicles have become an integral part of Intelligent Transport Systems (ITS) due to their ability to prevent hazardous traffic scenarios by deploying sensors and enabling vehicular communication for situational awareness. There are two main contenders of vehicular communication technologies: the Dedicated Short-Range Communication (DSRC) and Cellular-Vehicle-to-Everything (C-V2X). Until recently, DSRC, which is based on the IEEE 802.11p standard \cite{jkenney:dsrcmain}, was of the highest interest among researchers. 
However, with the emergence of 5G, regulators have taken a keen interest in the C-V2X technology which was standardized by the 3rd Generation Partnership Project (3GPP) Release 14 \cite{3gpp:3gppmain}. 



\begin{figure}[htbp]
\centerline{\includegraphics[trim=10 35 25 0,clip,width=.495\textwidth]{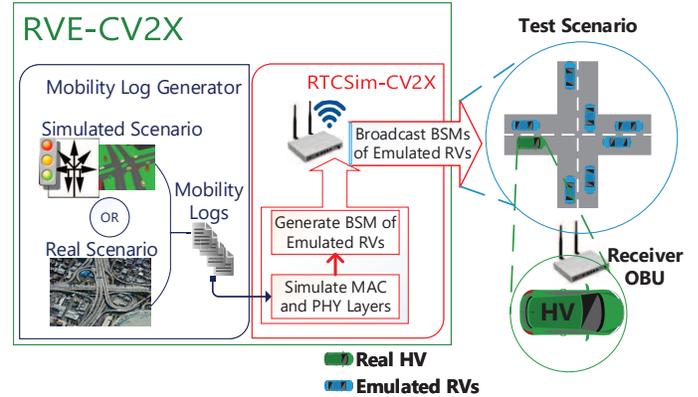}}
\caption{RVE-CV2X Framework}
\label{fig1}
\end{figure}

Despite an increasing interest in C-V2X,  
it is neither economically nor physically feasible 
to arrange hundreds of real Vehicular User Equipments (VUEs) for various testing scenarios to assess the performance and safety of C-V2X before its mass-deployment.
A few tools have been recently developed that can model these tests without relying on real VUEs. However, none of these tools can provide even close to real-time results, 
and thus cannot be utilized for real-time testing and emulation of large-scale scenarios.
Therefore, in this paper, the authors propose a
real-time, high-fidelity, and hardware-in-the-loop (HIL) emulation framework for C-V2X based connected vehicle applications (RVE-CV2X). 
RVE-CV2X can emulate any large-scale test scenario with any desirable configuration of road topologies, channel models and traffic level, without losing accuracy of a full-stack C-V2X communication system simulator.

The proposed RVE-CV2X system as shown in figure \ref{fig1} mainly consists of a mobility log generator tool, a Real-Time Communication Simulator (RTCSim-CV2X) and C-V2X on-board units (OBUs). The mobility log generator is a scenario generation tool that provides vehicular movement data using either pre-recorded logs from real VUEs or  generated logs from simulated VUEs. The RTCSim-CV2X is then responsible for using these logs to emulate the MAC and PHY layer behavior of the VUEs from the viewpoint of the host vehicle ($HV$). The ability of RTCSim-CV2X to achieve real-time emulation is the major enabling factor of this design. Throughout this emulation, the radio spectral behavior is broadcasted via the attached 
CV2X OBU in a real-time fashion. The $HV$ interacts with this broadcasted signal at its own CV2X OBU's transceiver as it would in a real test scenario. 


\section{Research Motivation}

The main motivation behind this study is to address the unfeasibility of large-scale testing using real VUEs and the run-time issue associated with the currently used simulators.
Testing of a technology 
is one of the most crucial steps before its deployment, specially when user safety is of prime concern.
However, in the case of C-V2X, 
it is unfeasible to perform large-scale testing by physically arranging hundreds of real VUEs for different testing scenarios solely to ensure an optimized network communication. Additionally, there can be tests involving near-collision or collision scenarios which can prove to be quite dangerous to be physically performed on a large scale. Repeatability is also another major issue in real-life field testing since it is extremely difficult to recreate a physical testing scenario in the exact possible way as in a previously executed test case.

Given the challenges associated with physical field testing, simulation frameworks are widely used as an alternative for testing the performance of vehicular communication.
There are a few network simulators based on frameworks such as NS-3 and OMNET++ that are currently being used to model C-V2X communication.
The study in \cite{Javier:syslevel} uses a simulator based on Viens simulation platform that is integrated with OMNET++ to evaluate the system-level performance of C-V2X mode 4 for an urban scenario defined in \cite{3gpp:studylte} under realistic traffic conditions. The paper highlights the possible transmission errors that can arise with the usage of the distributed scheduling protocol used in C-V2X mode 4 known as sensing-based semi-persistent sensing (SB-SPS) \cite{3gpp:mainv14.8}, \cite{3gpp:macv14.8}.

 Authors of \cite{behrad:mult}, \cite{saifuddin2020performance}, \cite{congest}, \cite{spatio}, present a simulator based on NS-3 simulation framework to perform analysis of highly congested traffic scenarios. They provide an insight into the inter-dependence of different parameters of SB-SPS on the resource allocation procedure. 
 An open-source simulator using NS-3 is provided in \cite{Fabian:open} where the authors analyze the performance of C-V2X mode 4 by using a 3GPP reference Manhattan grid scenario and a worst-case scenario of 100m x 100m playground. There is another simulator presented in \cite{Amr:perf} based on NS-3 which analyzes the effect of SB-SPS parameters on the scheduling performance. 


Despite the potential of these 
simulators to accurately provide valuable insights into the system-level performance of C-V2X, these simulators are not 
suitable for use in real-time testing and emulation.
The main reason for that is because all of these simulation platforms are known to take several orders of magnitude higher run-time as compared to the actual duration of each test scenario. 
Real-time testing of a communication technology is 
the most 
effective way to test different vehicular safety features on board and to ensure the robustness of the system.
Hence, a design of an HIL emulator was envisioned in the case of DSRC communication in \cite{gshah:cavs} where a simulation for low, medium or high traffic density could be obtained in real-time and was validated against \cite{ypfallah:bookchapter}. 
The current study implements a similar HIL emulator framework for C-V2X by introducing novel abstractions specific to C-V2X mode 4 and allowing batch completion of processes.

The rest of the paper is organized as follows. Section III provides an overview of C-V2X technology and the distributed scheduling mechanism of SB-SPS. Section IV provides a detailed explanation of the main components of the proposed RVE-CV2X.
We provide simulation results of RTCSim-CV2X in section V and confirm its accuracy by comparing it to the NS-3 based C-V2X simulator in \cite{behrad:mult}. 
Finally, the paper is concluded in Section VI. 

\section{C-V2X Mode 4 Background}
\subsection{Frame Structure in C-V2X}
C-V2X utilizes Single-Carrier Frequency Division Multiple Access (SC-FDMA) and it supports channel bandwidths of both 10 MHz and 20 MHz. The physical channel is divided into smaller fragments in time and frequency domain. Fragments in the time domain are referred to in the literature as $frames$. Each $frame$ is equivalent to 10 ms and is further divided into 10 smaller fragments known as $subframes$. In the frequency domain, a $channel$ is divided into $subcarriers$ each of size 15 kHz. A $resource\ block$ (RB) consists of 12 $subcarriers$ or 180 kHz of frequency bandwidth and one time-slot, i.e. 0.5 ms. The resource scheduling granularity in C-V2X from the perspective of the physical layer is 1 ms or in other words, one RB-pair in time-domain. A VUE can utilize multiple RBs for transmission in a single $subframe$. Available RBs in a $subframe$ are considered to be divided into chunks known as $subchannels$. Thus, a packet transmission can occupy one or more $subchannels$ in a $subframe$ depending upon the packet size and the configured number of RBs in every $subchannel$. The first two enumerated RBs in every transmission constitute for the Sidelink Control Information (SCI).  


\subsection{Sensing-Based Semi-Persistent Scheduling}
C-V2X mode 4 adopts a distributed scheduling protocol to autonomously select radio resources for transmission. This protocol is known as SB-SPS \cite{3gpp:mainv14.8}, \cite{3gpp:macv14.8}. VUEs utilize SB-SPS to periodically reserve the selected $subchannels$ as Candidate Single-Subframe Resources (CSRs) 
which is determined by the Random Sidelink Resource Re-selection Counter (SLRRC). SLRRC is chosen randomly between 5 and 15 every time a VUE attempts the SB-SPS procedure. 
The SLRRC is decremented by 1 after every transmission until it becomes equivalent to 0. If the SLRRC of a VUE becomes 0 and it wishes to transmit another packet(s), it can continue to do so at the same CSR location with a resource re-selection probability (P\textsubscript{resel}) or it can choose to perform a new SB-SPS procedure with a probability (1-P\textsubscript{resel}). Based on the study in \cite{behrad:mult}, P\textsubscript{resel} is chosen to be 0.8 to attain optimum performance and to provide a VUE with a higher chance of reserving the same resource for a longer time instead of frequently performing SB-SPS.  

Let us consider a case where a VUE $V\textsubscript{M}$ needs to reserve a CSR at time $T$ by performing an SB-SPS procedure. Initially, it creates a selection window which constitutes for $subchannels$ between time $T$ and the preset latency (100 ms). $V\textsubscript{M}$ includes all the CSRs from the selection window into a set ($S\textsubscript{A}$). $V\textsubscript{M}$ also contains a sensing window where the physical layer stores the information received within all CSRs in the 1000 ms before time $T$. Based on information acquired from the sensing window, $V\textsubscript{M}$ excludes CSRs from $S\textsubscript{A}$ that fulfill at least one of the following two conditions:
\begin{itemize}
    \item $V\textsubscript{M}$ used a CSR for transmitting at a subframe within the sensing window. This implies that the corresponding subframe has not been monitored for any incoming information on that subframe.  
    \item $V\textsubscript{M}$ successfully received an SCI and its associating Transport Block (TB) at a subframe within the sensing window from another VUE $V\textsubscript{W}$ and the Reference Signal Received Power (RSRP) of the TB is higher than the configured SPS threshold. 
\end{itemize}

The above two steps are repeated with a 3 dBm increment of SPS threshold until the size of $S\textsubscript{A}$ is at least 20\% of the size of the selection window. $V\textsubscript{M}$ then creates a second set ($S\textsubscript{B}$) where it includes CSRs from $S\textsubscript{A}$ that experience the lowest average Received Signal Strength Indicator (RSSI) such that the size of $S\textsubscript{B}$ is exactly 20\% of the size of the selection window. Finally, one CSR is chosen at random from $S\textsubscript{B}$ and is considered for a number of transmissions.
The SB-SPS takes a significant amount of processing power and time in regular simulators such as the simulator in \cite{behrad:mult}. However, in this study, a novel approach is adopted to abstracticize this process with rather simpler but accurate calculations as discussed in the next section. 

\section{Proposed Solution}
%
The proposed RVE-CV2X is capable of incorporating different kinds of road scenarios (rural, freeway or urban) and traffic levels (low, medium or high). Any test scenario within RVE-CV2X involves a pre-configured number of VUEs. These VUEs can be divided up into a $HV$ and remote VUEs ($RVs$). $HV$ can be considered as the test VUE for the experiment and we use the reception of this VUE to measure the performance of RVE-CV2X. As explained in detail below, RVE-CV2X allows varying configurations of $RVs$ and $HV$ for any test. It is possible to utilize virtual $RVs$ and $HV$ based on microscopic traffic simulators such as Simulation of Urban Mobility (SUMO) \cite{behrisch2011sumo}. It is also possible to deploy virtual $RVs$ along with a real $HV$. RVE-CV2X emulates the behavior of all $RVs$ from the perspective of the $HV$. As a consequence, the $HV$ receives packets from the emulated $RVs$ and creates a map assuming that all the $RVs$ are in its actual proximity. 
The following subsections fully explain the major components of RVE-CV2X and the implementation details of RTCSim-CV2X. 

%
\subsection{System Architecture Design }
A high-level framework of RVE-CV2X is shown in figure \ref{fig1}. From a system-level perspective, RVE-CV2X is comprised of a mobility log generator, RTCSim-CV2X, and C-V2X OBUs as detailed below.
%
\\

\subsubsection{Mobility Log Generator}
Mobility log generator is the initial component of RVE-CV2X.
It is run on a PC and is responsible for generating mobility logs for a pre-configured number of VUEs based on the specified road and traffic scenarios.
Each VUE's log file contains map-sharing information at each time-stamp such as vehicle identification, velocity, acceleration, latitude, longitude, heading, braking speed, and other relevant information.
As observable in figure \ref{fig1}, the mobility log generator can generate these logs either through real or simulated VUEs or through a combination of both.

For instance, if a user prefers to use virtual VUEs for the $HV$ and $RVs$, the mobility log generator can generate simulation-based mobility trace files using SUMO. 
At first,  the OpenStreetMap API can be used to feed a chosen map to SUMO. The user can then configure the desired traffic scenario and begin the SUMO simulation for the desired duration. The user can choose any specific VUE among the total number of VUEs to represent the $HV$ and the remaining VUEs as $RVs$. 


On the other hand, if a user prefers to generate a scenario with virtual VUEs to represent $RVs$ and an actual VUE for $HV$, the user can still utilize SUMO to generate logs for the virtual $RVs$. Next, the user can drive the real $HV$ within the same map boundary as used for SUMO simulation for the specified amount of time while the HIL is available as an on-board setup. As the $HV$ is driven, the RVE-CV2X gets live feed from the $HV$ mobility and uses it inside the RTCSim-CV2X for real-time emulation. In order to avoid any possible synchronization or VUE-on-VUE problems, the $HV$ can follow a loosely pre-defined path within the map boundary.

\bigskip
\subsubsection{RTCSim-CV2X}
RTCSim-CV2X is the second and the most significant component of RVE-CV2X.
RTCSim-CV2X takes mobility trace files generated by the mobility log generator as input and then emulates the C-V2X system, specifically the MAC and PHY layer behavior of the emulated $RVs$ from the perspective of the $HV$.
The RTCSim-CV2X module was initially allocated to run on a C-V2X on-board unit, however, due to the memory and processor limitations associated with the available on-board units, RTCSim-CV2X is required to be deployed on a PC. For this study, RTCSim-CV2X is run on a Linux PC with Core i7-6700 processor and 32 GB of RAM.
The error model implemented for RTCSim-CV2X in this study is the NIST error model which is the same as in the configuration of \cite{behrad:mult}.

%

%
%

\bigskip
\subsubsection{C-V2X OBUs}

Throughout the simulation of RTCSim-CV2X, whenever a packet such as Basic Safety Message (BSM) 
is supposed to be broadcasted, it is first transmitted from the PC via an ethernet cable to a C-V2X transceiving OBU within the RTCSim-CV2X block as shown in figure \ref{fig1}. The OBU then broadcasts the BSM packet over-the-air which is received by the C-V2X OBU connected to the $HV$ based on the emulated distance and channel congestion.

\subsection{Implementation Details}
\begin{figure}[t]
\centerline{\includegraphics[trim=20 0 20 0,clip,width=.48\textwidth]{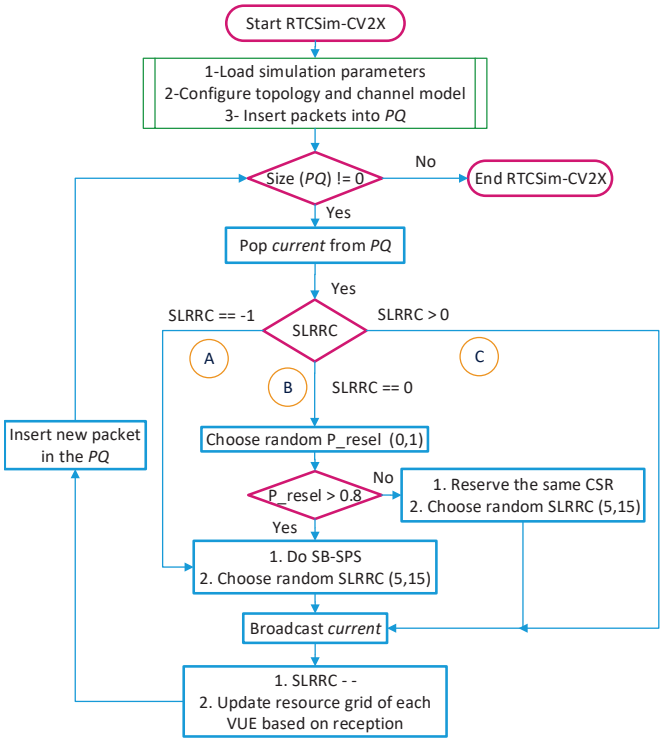}}
\caption{RTCSim-CV2X Flowchart}
\label{fig2}
\end{figure}

This sub-section explains the implementation details of RTCSim-CV2X. The flow-chart representation of the implemented logic of the RTCSim-CV2X module is shown in figure \ref{fig2}. At first, all the simulation parameters as shown in table \ref{table:configs} are set up before initializing the simulation. As simulation proceeds, the initial log entry from each VUE's mobility trace file generated by the mobility log generator is converted into a custom BSM packet and pushed inside a priority queue ($PQ$) which is sorted based on the packet generation time-stamps. This ensures that the packet that has the earliest packet generation time-stamp ($current$) will always be at the top of the queue. $current$ is popped from $PQ$ and based on the SLRRC value of the VUE containing the $current$ packet, the algorithm goes into one of the three main conditions, namely $A$, $B$ and $C$. 

The simulator continues to operate until log entries belonging to all the mobility trace files of all VUEs have been processed and thus the $PQ$ becomes empty. 
After each BSM packet transmission through one of the three conditions, 
a new log entry from the mobility trace file of the $current$ packet's VUE is customized into a BSM format, pushed into the $PQ$ and sorted based on its timestamp. The afore-mentioned conditions $A$, $B$ and $C$ are explained in detail below.    
\\
\subsubsection{A. SLRRC == -1}

Whenever RTCSim-CV2X goes into this condition, it implies that this is the first transmission in the ongoing simulation run from the VUE representing the $current$ packet. The simulator performs SB-SPS procedure in order to transmit $current$ based on the SB-SPS parameters as shown in table \ref{table:configs}. Secondly, it also assigns a random SLRRC value between 5 and 15 to the VUE. The SLRRC is decremented by 1 after every periodic transmission. Once SB-SPS chooses a CSR for transmission, $current$ is broadcasted at the corresponding scheduled timestamp and the resource grids of all the receiving VUEs are updated based on successful reception.   
\\
\subsubsection{B. SLRRC == 0}
This condition in RTCSim-CV2X arises when the SLRRC of a VUE reaches 0 after a number of periodic transmissions. 
With a P\textsubscript{resel} of 0.8, the VUE chooses the same CSR for a new series of periodic transmissions with a new randomly chosen SLRRC. With a probability (1-P\textsubscript{resel}) of 0.2, the VUE chooses to perform a new SB-SPS procedure and select a new CSR location for a number of periodic transmissions equivalent to a new randomly chosen SLRRC.
\\
\subsubsection{C. SLRRC \textgreater \space 0}
RTCSim-CV2X proceeds to this condition whenever the SLRRC of $current$ is above zero. This implies that the VUE just needs to periodically transmit $current$ at the same corresponding location in the new selection window as in the last transmission. Based on the reception, all the receiving VUEs update their resource grids respectively and a new BSM packet from the data entry of the VUE is loaded in the $PQ$ and sorted. 
%

%
%

\subsection{Abstractions}
This subsection highlights some of the abstractions achieved in RTCSim-CV2X to reduce the run-time and provide a real-time emulation:


1. Based on the pre-configured simulation parameters and batch calculation, the RTCSim-CV2X is able to be abstracted to the level of $subchannels$ as opposed to RBs as done in \cite{behrad:mult}.  

2. All RBs within a $subchannel$ are abstracted to contain the same Power Spectral Density (PSD) value while the error model is modified to correctly introduce the power distribution as mentioned in [5]. The interference model also uses a statistical distribution of received power to reach the resultant average SINR for the BSM packets.


3. RTCSim-CV2X only focuses on the reception details of the $HV$. Although the $RVs$ are emulated in detail, significant run-time and memory space is saved by discarding reception workflow higher than MAC layers among the $RVs$. 
This is because the HIL requires the interference and transmission power from the perspective of the $HV$ only.

4. For the receiver model, transmission chunk processors used in NS-3 are absent in RTCSim-CV2X. Instead, calculation of SINR is performed in a periodic batch manner using derived behavior of receivers, which makes the step several orders of magnitude faster as compared to that of NS-3. 

In summary, the assumption of fixed simulation parameters and known system behavior is exploited to develop a managed code-base that outputs the same result as that of NS-3 but with much lesser calculations.

\section{Analysis and Results}
%
In this section, we have attempted to validate the accuracy of RTCSim-CV2X by comparing its performance to the NS-3 based C-V2X simulator used in \cite{behrad:mult}. Although NS-3 is widely used to accurately model V2X communication, it cannot provide real-time large-scale Vehicular Ad-Hoc Network (VANET) simulations. In fact, as shown later in this section, NS-3 based simulators are known to take several orders of magnitude higher run-time as compared to the actual simulated time. The simulation parameters used in this study for both NS-3 and RTCSim-CV2X to ensure a meaningful comparison are listed in table \ref{table:configs}. As mentioned earlier, these parameters can be easily reconfigured for 
a particular test scenario. 

\begin{figure}[t]
\centerline{\includegraphics[width=.48\textwidth]{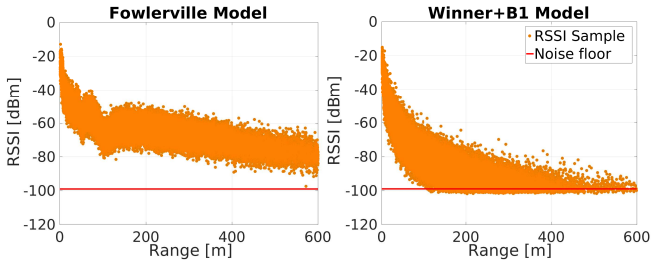}}
\caption{RSSI plots of Fowlerville and Winner+B1 Models}
\label{fig3}
\end{figure}

\begin{figure*}[t]
    \includegraphics[width=\textwidth,height=8cm]{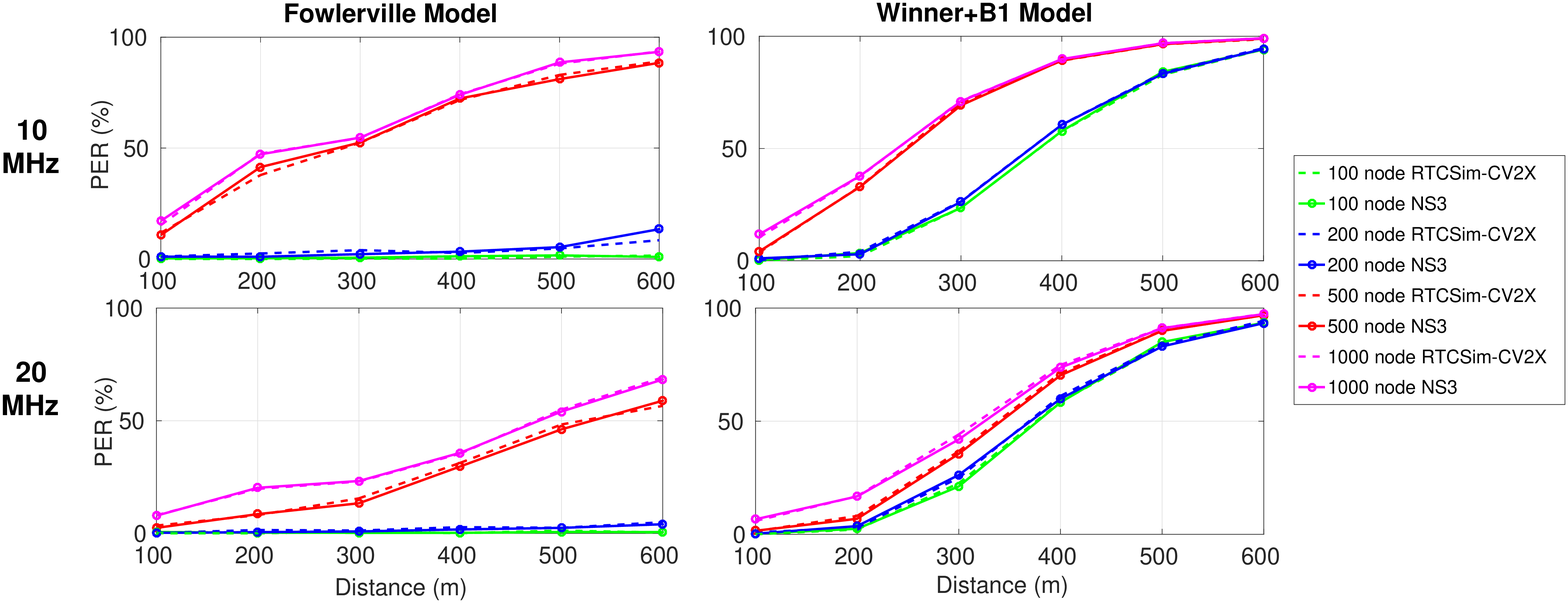}
    \caption{PER comparison of NS-3 and RTCSim-CV2X}
    \label{fig4}
\end{figure*}


%
\begin{table}[htbp]
\centering
\caption{SIMULATION PARAMETERS FOR NS-3 AND RTCSim-CV2X}
\begin{tabular}{l r}
\hline
\hline
Frequency Bandwidth(s)          & 10 \text{MHz}, 20 \text{MHz}\\
BSM Packet Size                  & 190 Bytes\\
Modulation and Coding Scheme      & 5\\
Packet Transmission Rate                 & 10 \text{Hz}\\
Transmission Power                         & 20 \text{dBm}\\
Resource Reservation Interval         & 100 \text{ms}\\
SLRRC                               & [5,15]\\
P\_resel    & 0.8\\
Simulation Time                       & $40s$\\
Error Model         & NIST\\
\hline
\end{tabular}
\label{table:configs}
\end{table}

The main performance metric used in this study for the comparison between NS-3 and RTCSim-CV2X is the Packet Error Rate (PER).
It is computed from the receiving VUE's perspective as the ratio between the number of non-decoded BSM packets and the total number of BSM packets being sent to the receiving VUE. 
Once the PER performance is confirmed to be accurate, RTCSim-CV2X is further analyzed to understand the timing dynamics of BSM packet receptions. This is done by utilizing the performance of Inter-Packet Gap (IPG). IPG is computed as the time difference between successfully decoded packets from a specific transmitter and received by a specific receiving VUE.



The road topology used for this experiment comprises of a 1200 m linear road with six lanes. To validate the results of RTCSim-CV2X for multiple traffic scenarios, we use the test cases of 100 (low density), 200 (medium density), 500 (high density) and 1000 (very high density) VUEs placed randomly on the afore-mentioned linear road with a constant velocity model. 
For each traffic scenario, there are two frequency bandwidths, i.e. 10 MHz and 20 MHz, and two channel models, i.e. Fowlerville freeway loss model \cite{camp2014interoperability},\cite{mahjoub2018composite} and Winner+B1 loss model \cite{bultitude20074}. Thus, in total there are 16 different test scenarios used for this study.
Fowlerville model is particularly useful when modeling freeway scenarios where it allows successful BSM reception even at higher distances. On the contrary, Winner+B1 model is usually utilized to model urban scenarios. It is common for urban scenarios to possess more traffic and high buildings which can drastically reduce the communication range of VUEs. Thus, Winner+B1 model provides successful reception only at short distances. 


Figure \ref{fig3} shows the RSSI plots of Fowlerville and Winner+B1 models for an example 500 VUE traffic scenario on the afore-mentioned topology to indicate the difference between these two channel models. The visual representation of RSSI scatter plots in figure \ref{fig3} and PER 
curves as shown later in this section, is limited to 600 m since it is critical to measure the performance of V2X communication, specially at shorter distances among VUEs. 
Figure \ref{fig3} exhibits a gradual reduction in the RSSI plot of Fowlerville model. It is also evident in the case of Fowlerville that the RSSI samples observed by the $HV$ are clearly above the noise floor even when the $RVs$ are 600 m apart from the $HV$. On the other hand, when Winner+B1 model is deployed, the RSSI plot can be seen to sharply fall as the distance between the $HV$ and $RVs$ is increased. In addition, numerous samples can be observed below the noise floor at distances of around 150 m and beyond.

\begin{figure*}[b]
    \includegraphics[trim=13 30 30 15,clip, width=\textwidth,height=8cm]{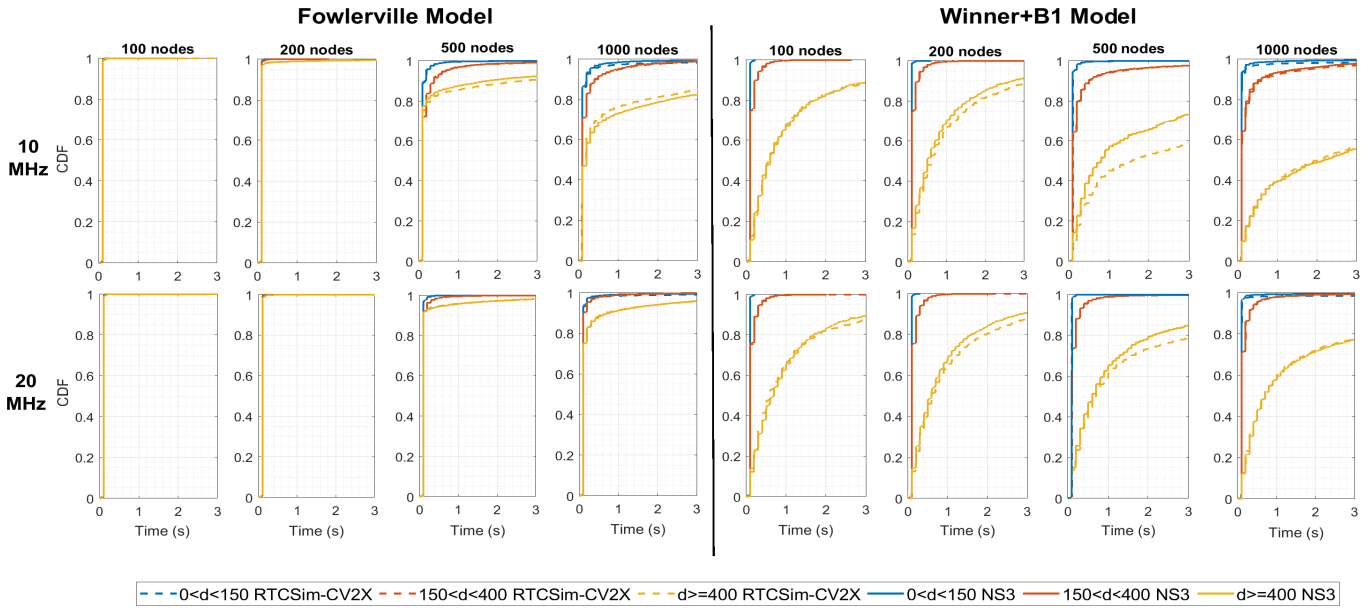}
    \caption{IPG comparison of NS-3 and RTCSim-CV2X}
    \label{fig5}
\end{figure*}

\begin{figure*}[t]
\centerline{\includegraphics[width=\textwidth,height=6cm]{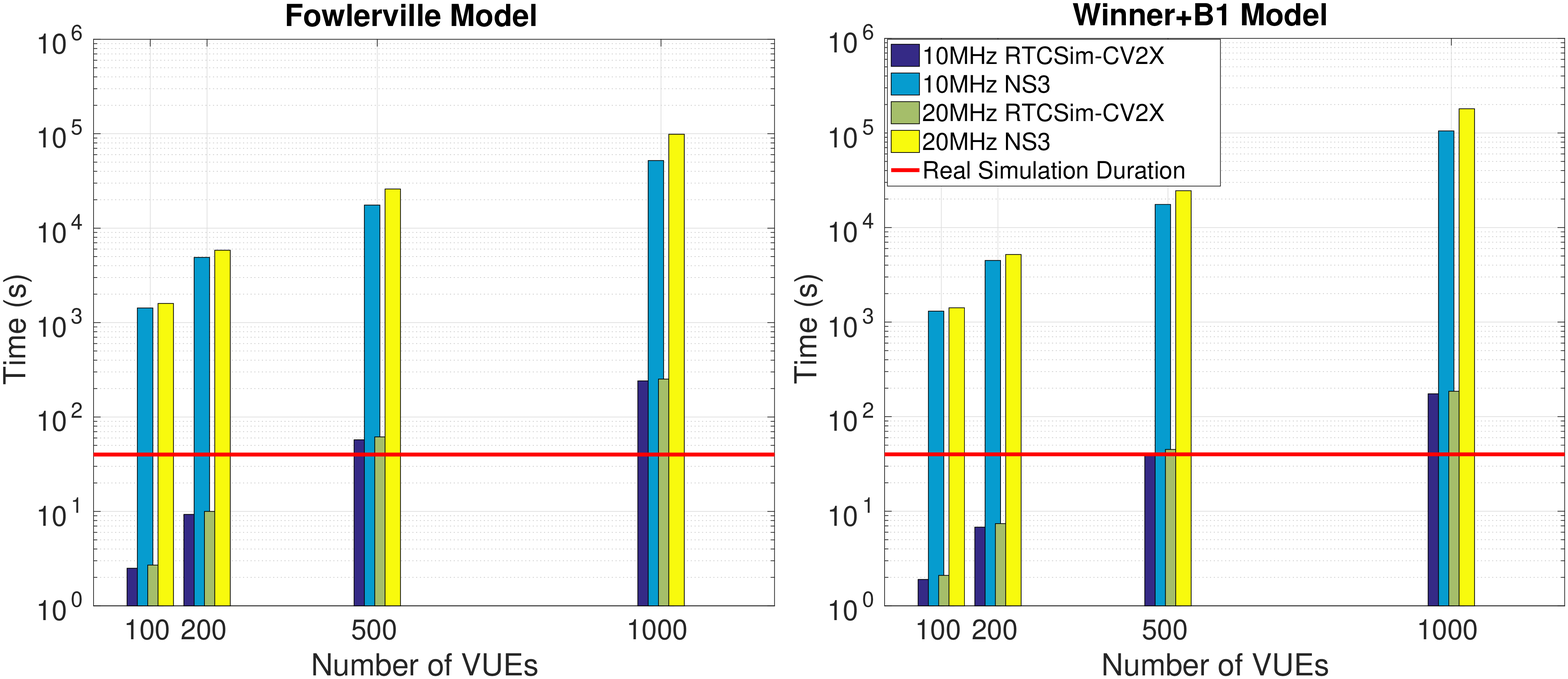}}
\caption{Run-time comparison of NS-3 and RTCSim-CV2X.}
\label{fig6}
\end{figure*}

Figure \ref{fig4} presents the PER comparison between NS-3 and RTCSim-CV2X for the earlier mentioned 16 test scenarios. The left sub-figures within figure \ref{fig4} show the PER results of 100, 200, 500 and 1000 VUE traffic scenarios using Fowlerville model with 10 MHz and 20 MHz bandwidths, respectively. On the other hand, the right sub-figures of figure \ref{fig4} present the PER plots of the same afore-mentioned traffic scenarios for 10 and 20 MHz channels but with Winner+B1 model.     
It is clearly evident from figure \ref{fig4} that the PER curves of RTCSim-CV2X are similar to those of NS-3 for all test scenarios. Generally, all PER plots tend to rise as the distance between $RVs$ and the $HV$ increases. It can also be seen that using wide-band (20 MHz) as opposed to narrow-band (10 MHz) makes a significant difference in the PER results for all scenarios. This can be explained by noticing that using 20 MHz channel roughly provides twice the number of CSRs to every VUE during SB-SPS as compared to using 10 MHz bandwidth. 
This eventually translates to a lower PER in the case of 20 MHz. 

For a given traffic scenario and frequency bandwidth, the difference in PER plots as portrayed in figure \ref{fig4} can also be associated with the difference in the performances of the two channel models. In general, PER can be seen to rise at a faster rate for Winner+B1 compared to that of Fowlerville. In the case of 100 and 200 VUE scenarios for both 10 MHz and 20 MHz, the PER plots for Fowlerville remain close to 0\% for all distance values between the $HV$ and $RVs$ up to 600 m. However, PER plots for the same scenarios can be observed to rise from 0\% at origin up to around 90\% at 600 m distance for Winner+B1. In the case of 500 and 1000 VUE scenarios where channel congestion is expected, Winner+B1 performs better than Fowlerville at close distances up to 200 m. However, PER for Winner+B1 rises sharply beyond 200 m and the plots for Winner+B1 using both 10 MHz and 20 MHz reach almost 100\% PER at 600 m, whereas the plots for Fowlerville reach around 90\% PER in the case of 10 MHz and 60\% PER in the case of 20 MHz.

Since the PER curves of RTCSim-CV2X are similar to those of NS-3, the timing dynamics of the received packets can now be compared as well. For that, the performance of IPG results of RTCSim-CV2X is compared to that of NS-3 for all 16 test scenarios as shown in figure \ref{fig5}. Three main distance bins are used for the IPG plots: a critically low distance range from 0 m to 150 m, a medium distance range from 150 m to 400 m, and a high distance range greater than 400 m. Primary focus is placed on the low and medium distance bins since they are critical from the perspective of any safety application.
It can be clearly observed that IPG results of RTCSim-CV2X for low and medium distance bins are extremely similar to those of NS-3 for all 16 test scenarios. This exemplifies that the SB-SPS procedure within RTCSim-CV2X is performing in an  identical fashion as that within NS-3. Even for bins corresponding to high distances greater than 400 m, the IPG performance of RTCSim-CV2X is similar to that of NS-3. 

It can be noticed that only one particular scenario involving 500 VUEs in figure \ref{fig5} with Winner+B1 channel model and 10 MHz bandwidth shows a difference between RTCSim-CV2X and NS-3 for IPG plots beyond 400 m although PER plots for the same scenario in figure \ref{fig4} are similar. 
An analysis of this exact scenario by detailing the output of each module in the RTCsim-CV2X workflow confirms that
this deviation mainly arises due to the receiver model of RTCSim-CV2X which follows a probabilistic calculation. 
For the afore-mentioned scenario,
the variation in IPG results for higher distance bins occurs
where packet reception falls below 10\% and IPG becomes sensitive to the reception order at each periodic resource. 
On the contrary, this behavior is not observed in the scenario with 1000 VUEs for the same channel and bandwidth as shown in figure \ref{fig5} because the average collision distance among VUEs is smaller as compared to the scenario with 500 VUEs and thus the receiver model is able to provide very similar results to that of NS-3. This difference in IPG results is also not duplicated in scenarios comprising of lower number of VUEs (100 and 200 VUEs) 
due to scarcity of packet collisions within the reception range. 
Therefore, an effort to further improve the receiver model is postponed as that will contribute little to the actual performance measurement of the HIL where the longer distance communication has lesser importance.   
 


In addition to the validation of RTCSim-CV2X through PER and IPG, another significant benefit of this simulator is its capability to run identical scenarios as NS-3 at a considerably lower run-time. Figure \ref{fig6} shows the run-time comparison of RTCSim-CV2X and NS-3 and it is evident that RTCSim-CV2X is several orders of magnitude faster than NS-3 for all test scenarios. The plots in figure \ref{fig6} also include a horizontal line to reflect the actual simulation duration of all the example test scenarios, i.e. 40 s. 
It can be visualized that the NS-3 run-time is considerably higher than the simulation duration for all test scenarios. Table \ref{table:ExecutionTime} also provides a tabular version of average run-time comparison of every test scenario and reflects a vast difference in the run-time of RTCSim-CV2X and NS-3. It should be noted that based on the hardware utilized for this study, RTCSim-CV2X has a current capability of providing real-time simulation for up to around 500 VUEs with a maximum resource consumption of 988MB.

\begin{table}[t]
\centering
\caption{COMPARISON OF MEAN EXECUTION TIME OF NS-3 AND RTCSim-CV2X}
\begin{center}
\begin{tabular}{|c||c||c|c|c|c|}
\hline
\multicolumn{6}{|c|}{MEAN EXECUTION TIME (s) for 40 s SIMULATION}                                                                \\ \hline
\multirow{2}{*}{Channel Model} & \multirow{2}{*}{VUEs} & \multicolumn{2}{c|}{\textit{10 MHz}} & \multicolumn{2}{c|}{\textit{20 MHz}} \\ \cline{3-6} 
                                &       & \textbf{RTCSim}   & NS-3      & \textbf{RTCSim}       & NS-3      \\ \hline \hline
\multirow{3}{*}{Fowlerville}           & 100   & \textbf{ 2.5}   & 1428      & \textbf{ 2.7}       & 1594     \\ \cline{2-6} 
                                & 200   & \textbf{ 9.3}   & 4899     & \textbf{ 10.0}       & 5835     \\ \cline{2-6} 
                                & 500  & \textbf{ 57.4}   & 17545     & \textbf{ 61.5}       & 26010     \\ \cline{2-6}
                                & 1000  & \textbf{ 241.1}   & 52002     & \textbf{ 252.0}       & 98860     \\ \hline \hline
\multirow{3}{*}{Winner+B1}         & 100   & \textbf{ 1.9}   & 1308     & \textbf{ 2.1}       & 1417     \\ \cline{2-6} 
                                & 200   & \textbf{ 6.8}   & 4487     & \textbf{ 7.4}       & 5185     \\ \cline{2-6} 
                                & 500  & \textbf{ 40.5}   & 17563     & \textbf{ 45.0}       & 24499     \\ \cline{2-6}
                                & 1000  & \textbf{ 174.6}   & 105116     & \textbf{ 185.5}       & 180149     \\ \hline \hline
\end{tabular}
\end{center}
\label{table:ExecutionTime}
\end{table}

\section{Concluding Remarks}
%

This study presents RVE-CV2X, which is a real-time, high-fidelity, and HIL emulation framework for C-V2X based VANET applications.
The accuracy of RTCSim-CV2X within RVE-CV2X is validated for various road and traffic scenarios by comparing its PER and IPG performance with an already validated simulator based on the NS-3 framework. It is also shown that RTCSim-CV2X is several orders of magnitude faster than the NS-3 based simulator.
Although the framework focuses only on the reception details of the $HV$, the real-time nature of this tool allows the user to emulate a scenario for enough duration without any spatial constraint, thereby enabling the system performance to be determined for any traffic situation.



In addition to the current contributions mentioned in this paper, there can be further improvements and use-cases of RVE-CV2X. 
Currently, only a few aspects of the SB-SPS procedure within RVE-CV2X have been abstracted and many of its functions are elaborate implementations of the MAC procedures. However, it is possible to introduce more abstractions by using a probabilistic theoretical model of SB-SPS once it becomes available. 
Secondly, due to the modular nature of RTCSim-CV2X, the current design can be modified to allow input from any additional application in addition to basic safety that uses shared radio resources, i.e. tolling, infotainment, etc. 
Finally, with further research, the scalability for real-time performance of RVE-CV2X can be improved by allowing parallel emulation of independent subsets of procedures using additional hardware resources. 

%
\balance
\bibliography{main.bib}{}
\bibliographystyle{unsrt}
\end{document}